\DocumentMetadata{}
\documentclass[sigconf]{acmart}

\renewcommand\footnotetextcopyrightpermission[1]{} 

\usepackage{pifont}
\usepackage{amsmath}
\usepackage{amsfonts}
\usepackage{graphicx}
\usepackage{textcomp}
\usepackage{xcolor}
\usepackage{multirow}
\usepackage{soul}
\usepackage{tikz}
\usepackage{booktabs}
\usepackage{algorithm}
\usepackage{algpseudocode}
\usepackage{colortbl}
\usepackage{bm}
\usepackage{float}
\usepackage{algorithm}
\newcommand\encircle[1]{%
\tikz[baseline=(X.base)] 
  \node (X) [draw, scale=0.75, shape=circle, inner sep=0, fill=black, text=white, minimum size=0em] {\strut #1};}

\usepackage{amsmath,amsfonts,bm}

\def\eqref#1{equation~\ref{#1}}

\def\1{\bm{1}}

\def\vt{{\bm{t}}}

\def\vx{{\bm{x}}}

\DeclareMathAlphabet{\mathsfit}{\encodingdefault}{\sfdefault}{m}{sl}
\SetMathAlphabet{\mathsfit}{bold}{\encodingdefault}{\sfdefault}{bx}{n}
\newcommand{\tens}[1]{\bm{\mathsfit{#1}}}

\def\tB{{\tens{B}}}

\begin{document}

\title{DNN-Defender: A Victim-Focused In-DRAM Defense Mechanism for Taming Adversarial Weight Attack on DNNs}
\author{Ranyang Zhou$^{\dagger,*}$, Sabbir Ahmed$^{\ddagger,*}$, Adnan Siraj Rakin$^\ddagger$, and Shaahin Angizi$^\dagger$}
\affiliation{
\institution{\small$^\dagger$Department of Electrical and Computer Engineering, New Jersey Institute of Technology, Newark, NJ \country{USA}}{}
\institution{\small$^\ddagger$Department of Computer Science, State University of New York at Binghamton, NY \country{USA}}{}
\institution{$^*$These authors contributed equally}{}}
\email{rz26@njit.edu, sahmed9@binghamton.edu, arakin@binghamton.edu, shaahin.angizi@njit.edu} \vspace{-1em}

\begin{abstract}
With deep learning deployed in many security-sensitive areas, machine learning security is becoming progressively important. Recent studies demonstrate attackers can exploit system-level techniques exploiting the RowHammer vulnerability of DRAM to deterministically and precisely flip bits in Deep Neural Networks (DNN) model weights to affect inference accuracy. The existing defense mechanisms are software-based, such as weight reconstruction requiring expensive training overhead or performance degradation. On the other hand, generic hardware-based victim-/aggressor-focused mechanisms impose expensive hardware overheads and preserve the spatial connection between victim and aggressor rows.
In this paper, we present the first DRAM-based victim-focused defense mechanism tailored for quantized DNNs, named \textit{DNN-Defender} that leverages the potential of in-DRAM swapping to withstand the targeted bit-flip attacks with a priority protection mechanism. Our results indicate that DNN-Defender can deliver a high level of protection downgrading the performance of targeted RowHammer attacks to a random attack level. In addition, the proposed defense has no accuracy drop on CIFAR-10 and ImageNet datasets without requiring any software training or incurring hardware overhead.
\end{abstract}\vspace{-6em}




\maketitle
\pagestyle{plain} 

\section{Introduction}
\begin{figure}[t]
\begin{center}
\begin{tabular}{c}
\includegraphics [width=0.97\linewidth]{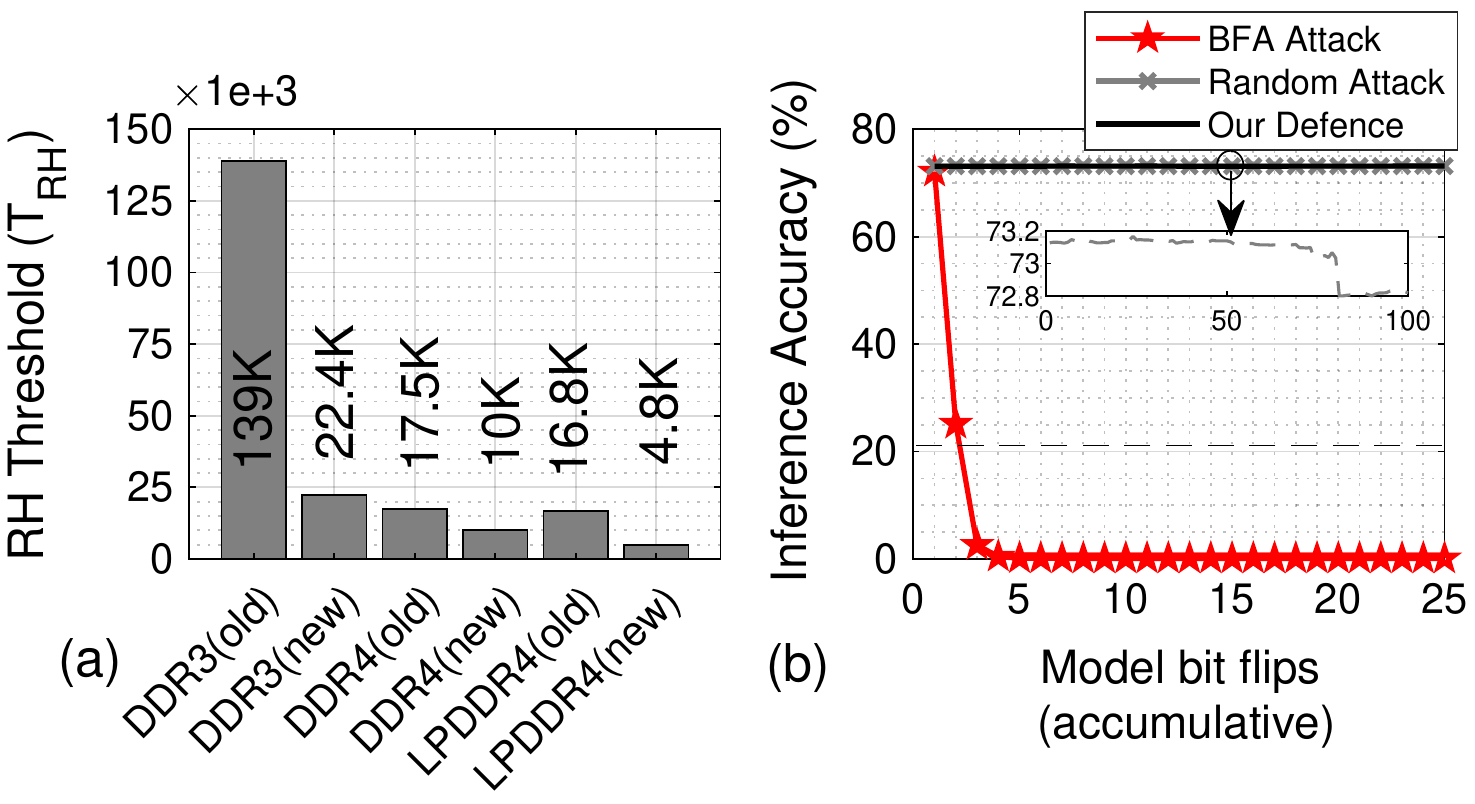}\vspace{-0.4em}
 \end{tabular} \vspace{-1.2em}
\caption{(a) RowHammer thresholds \cite{woo2022scalable}, (b) Targeted vs. random bit flipping for an 8-bit ResNet-34 on ImageNet and \textit{DNN-Defender}'s performance.}\vspace{-2.2em}
\label{RHthre}
\end{center}
\end{figure}

The far-reaching development of Deep Neural Network (DNN) accuracy even with low-bit-width models has recently triggered various security-associated attacks in many applications \cite{rakin2019bit}. Recent studies show that an adversary can identify and manipulate a small number of vulnerable bits of off-the-shelf well-trained DNN weight parameters to significantly compromise the output accuracy \cite{hong2019terminal,rakin2019bit,ahmed2024deep}. Such Bit-Flip Attacks (BFAs) have been enabled mainly due to a manifestation of a DRAM cell-to-cell interference and failure mechanism called RowHammer (RH) \cite{kim2014flipping}. RH attack is conducted when a malicious process activates and pre-charges a specific row (i.e., aggressor row) repeatedly to a certain threshold ($T_{RH}$) to induce bit-flips on immediate nearby rows (i.e., victim rows). 
Unfortunately, by scaling down the size of DRAM chips in the modern manufacturing process, DRAM becomes increasingly more vulnerable to RH bit-flips \cite{kim2020revisiting}. Figure \ref{RHthre}(a) shows that the RH threshold has had a significant downward trend in recent years, e.g., the attacker needs $\sim$4.5$\times$ fewer hammer counts on LPDDR4 (new) as opposed to DDR3 (new) \cite{woo2022scalable}.

To prevent RH attacks, DRAM manufacturers and researchers have proposed hardware-based victim-focused defense mechanisms to proactively refresh the victim rows by adding counters to count the number of activations \cite{kim2014flipping,zhou2023threshold}. However, such RH mitigation proposals have faced a huge overhead both from latency and power consumption perspectives \cite{zhou2022lt}. To mitigate this, recent aggressor-focused swap-based mechanisms \cite{saileshwar2022randomized,woo2022scalable} proactively swap and unswap aggressors with random rows before reaching the RH threshold. Such mechanisms can be immensely effective when the attacker does not have knowledge of the internal DRAM organization. 
The Randomized Row-Swap (RRS) \cite{saileshwar2022randomized} swaps the aggressor row with a random row within the same bank in the memory. A method called Secure Row-Swap (SRS) \cite{woo2022scalable} has demonstrated the use of fewer counters for crucial data and implemented associated threat mitigation using the swap operation. This approach reduces the swap rate while maintaining security, resulting in higher efficiency and lower latency. 
Assuming the attacker’s access
to this information, the attacker will not track the aggressor row but the victim row and attack its adjacent row, making it a new aggressor row. In this case, swapping the aggressor row with another random row is purposeless. On the other side, prior works have approached the problem of DNN weight noise mitigation from a software-training optimization perspective \cite{he2020defending,li2020defending} and DNN architecture modification~\cite{rakin2021ra} that impose expensive overhead or performance/accuracy degradation.

We develop DNN-Defender as a pure DRAM-based victim-focused defense mechanism to effectively withstand the targeted RH BFAs on DNNs. The main contributions of this work are:
(1) We design the DNN-Defender mechanism with hardware-software support that utilizes in-DRAM swapping to protect the DNN weight parameters not requiring any software training or imposing additional hardware overhead; 
(2) We develop a priority protection mechanism method and parallelism to tailor the performance-accuracy trade-offs with respect to the system requirements; and
(3) We extensively analyze the DNN-Defender's applicability and efficiency in taming RH vulnerability compared to recent hardware/software techniques over CIFAR-10 and ImageNet DNN datasets. On the CIFAR-10, the accuracy of the DNN-Defender-supported system is 91.71\% while the baseline indicates only 10.9\% under BFA.\vspace{-0.5em}

\section{Background \& Motivation} \vspace{-0.5em}
\subsection{DRAM}
\begin{figure}[t]
\begin{center}
\begin{tabular}{c}
\includegraphics [width=0.97\linewidth]{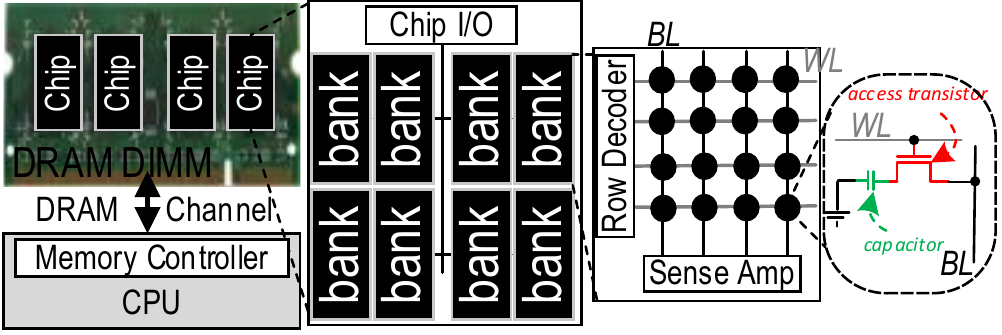}\vspace{-0.4em}
 \end{tabular} \vspace{-0.8em}
\caption{Organization of a DRAM chip.}\vspace{-1em}
\label{DRAM} \vspace{-1em}
\end{center}
\end{figure}

\vspace{-0.5em}
\textbf{Organization \& Commands.} The DRAM chip is a hierarchical structure consisting of several memory banks as shown in Fig. \ref{DRAM}. Each bank comprises 2D sub-arrays of memory bit-cells that are virtually ordered in memory matrices (mats), which have billions of DRAM cells on modern chips. Each DRAM bit-cell consists of a capacitor and an access transistor. The charge status of the bit cell's capacitor is used to represent binary ``1'' or ``0'' \cite{seshadri2017ambit,angizi2019graphide}. In idle mode, the memory controller turns off all enabled DRAM rows by sending the Precharge (PRE) command on the command bus. This will precharge the Bit-Line (BL) voltage to $\frac{V_{DD}}{2}$. In the active mode, the memory controller will send an Activate (ACT) command to the DRAM module to activate the Word-Line (WL). Then, all DRAM cells connected to the WL share their charges with the corresponding BL. Through this process, BL voltage deviates from the precharged $\frac{V_{DD}}{2}$. The sense amplifier then senses this deviation and amplifies it to $V_{DD}$ or 0 in the row buffer. The memory controller can then send read (RD)/write (WR) commands to transfer data to/from the sense amplifier array \cite{zhou2022red,zhou2022flexidram}.

\textbf{RowClone.}
Exploiting the fact that DRAM transfers an entire row of data to the corresponding row buffer during the read operation, RowClone \cite{seshadri2013rowclone} has been developed as a simple and efficient mechanism to enable a bulk in-memory copy operation ($<$100ns) from a source row to a destination row completely in the DRAM sub-array. RowClone eliminates the need to transfer data over the memory channel. The memory controller manages this by issuing two back-to-back ACT commands first to the source and then the destination without a PRE command in between with almost negligible cost. By using this method, the latency and power consumption of a bulk copy operation can be reduced by a factor of 11.6 and 74.4, respectively \cite{seshadri2013rowclone}.\vspace{-1em}

\subsection{Row Hammer-based DNN Weight Attack}
The BFA progressively searches for vulnerable bits by first performing a bit ranking within each layer based on gradient \cite{rakin2019bit}. Considering a weight quantized DNN, the weight matrix can be parameterized by two complement representations $\{\tB_l\}_{l=1}^L$, where $l \in \{1, 2,...,L\}$ is the layer index. 
BFA computes the gradient w.r.t. each bit of the model ($ |\nabla_{\tB_l} \mathcal{L}|$) where $\mathcal{L}$ is the inference loss function. At each iteration, the attacker performs two key attack steps: i) inter-layer search and ii) intra-layer search; where the goal is to identify a vulnerable weight bit and flip it. Given a sample input $x$ and label $t$, the BFA~\cite{rakin2019bit} algorithm tries to maximize the following loss function ($\mathcal{L}$):
\vspace{-8pt}
\begin{equation}
\label{eqt:BFAN}
\begin{gathered}
\max_{\{\hat{\tB}_l\}}  ~\mathcal{L}\Big (f \big( \vx ; \{\hat{\tB}_l\}_{l=1}^{L} \big), {\vt} \Big),
\end{gathered}
\vspace{-5pt}
\end{equation}

\noindent while ensuring the hamming distance between the perturbed weight tensor by BFA ($\hat{\tB}_{l=1}^L$) and initial weight tensor ($\{\tB_l\}_{l=1}^L$) remains minimum. Finally, the attack efficiency can be measured by the number of bit-flips required to cause DNN malfunction. 
\begin{figure}[t]
\begin{center}
\begin{tabular}{c}
\includegraphics [width=0.92\linewidth]{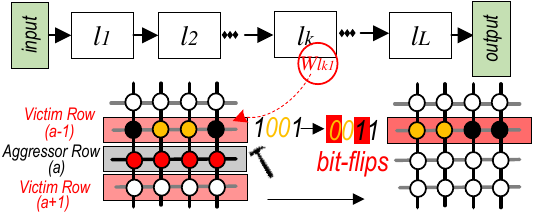}\vspace{-0.4em}
 \end{tabular} \vspace{-0.8em}
\caption{Adversarial weight RowHammer attack in the $k_{th}$ layer of a DNN.}\vspace{-1em}
\label{RH} \vspace{-0.5em}
\end{center}
\end{figure}
Figure \ref{RHthre}(b) illustrates how the DNN accuracy degrades under a few (less than 5) targeted bit-flips using DeepHammer attack \cite{yao2020deephammer} in an 8-bit quantized ResNet-34 running the ImageNet dataset as opposed to over 100 random BFAs.
Figure \ref{RH} illustrates how such an adversarial weight attack is conducted in an $L$-layer DNN on the target $W_{l_{k1}}$ = 1001, i.e., the weight located in the $l_k$ layer. The malicious process continuously hammers the aggressor row in $a$-address and induces bit-flips on adjacent victim rows ($a-1$ and $a+1$) holding $W_{l_{k1}}$. As a result, such a single-sided RH attack changes the weight value (herein,  1001$\rightarrow$\ul{0}0\ul{1}1).\vspace{-0.5em}

\section{White Box Threat Models}
\begin{figure}[b] \vspace{-1.5em}
\begin{center}
\begin{tabular}{c}
\includegraphics [width=0.82\linewidth]{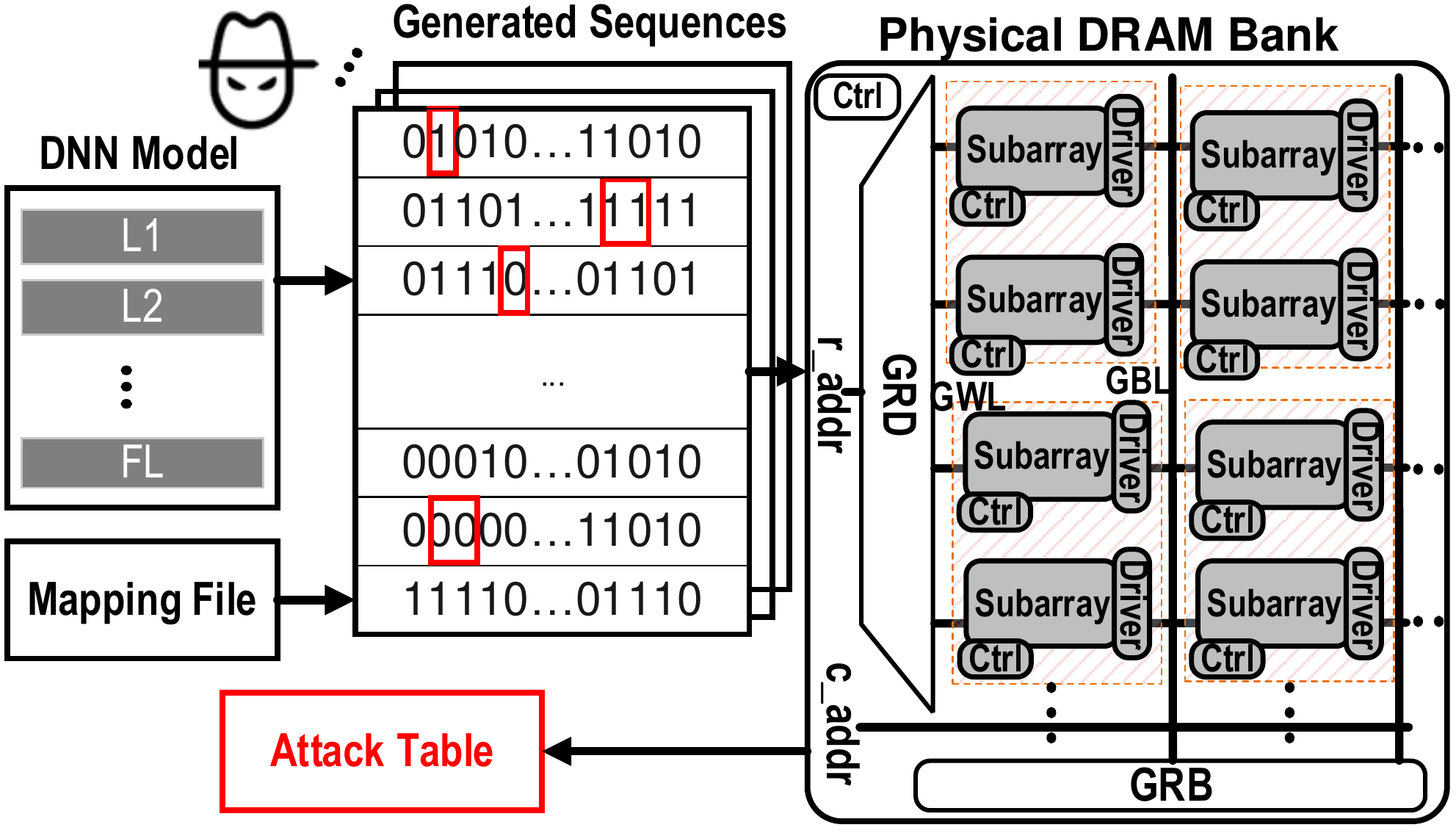}
 \end{tabular} \vspace{-0.8em}
\caption{Threat model about attacker’s known information about DNN and the attack process.}\vspace{-2em}
\label{model}
\end{center}
\end{figure}

\textbf{Hardware Threat Model.}
We assume the following threat model: 1) Each row has a threshold $T_{RH}$ after becoming an aggressor row, and once exceeded within the refresh interval ($T_{ref}$), it will impose a bit-flip to two adjacent victim rows; 2) We assume that all vulnerable data rows are neither concentrated in one/two sub-arrays nor evenly distributed in each sub-array. Experimentally, most sub-arrays store several data rows simultaneously; some may store multiple or none; and 3) The attacker has a detailed mapping file as shown in Fig. \ref{model} that can locate the physical address of the target data in the neural network and is aware of the initial static mapping of the DRAM rows (i.e., physical adjacency information between rows) 
\cite{jattke2022blacksmith,wi2023shadow}. Therefore, the attacker can perform an RH attack on the targeted content.

\textbf{Software Threat Model.}
We assume a standard white-box threat model for the BFA adopted across multiple attack domains in prior works~\cite{rakin2019bit,rakin2021t}. In a white-box threat model (summarized in Table \ref{tab:threat}), the attacker is aware of the internal structure of the DNN models, e.g., the number of layers and the width of each layer. On top of that, the attacker has complete knowledge of the DNN model parameters, their values, and bit representation for inference. This assumption is practical due to the recent advancement of side-channel information leakage and the reverse-engineering of DNN models~\cite{yan2020cache} that can recover the DNN model configuration at the inference stage.
However, the attacker can not access the training stage configuration and data even in a white-box setting. The attacker has a sample batch of test data to launch the attack. Moreover, the attacker is co-located with the victim, which enables the attacker to run user-space unprivileged processes. Finally, the attacker has complete knowledge of the DRAM addressing scheme. 
In this work, two types of white-box attack threat models are considered. A \textit{\ul{semi-white-box attack}}, as a weak version of the BFA, where the attacker is unaware of our proposed defense scheme, and a \textit{\ul{complete white-box attack}}, where the attacker is aware of the defense and takes necessary actions to circumvent the mitigation. \vspace{-0.4em}

\begin{table}[t]
\centering
\caption{Standard threat model for BFA~\cite{rakin2019bit,yao2020deephammer}.}
\label{tab:threat} \vspace{-1em}
\scalebox{0.7}{
\begin{tabular}{cc}
\hline
 Information & Attackers Access \\ \hline
 Model architecture and parameters & $\checkmark$ \\
 Small batch (e.g., 128) of test data & $\checkmark$ \\
 Address of parameter cached in DRAM & $\checkmark$ \\
  Model training data and configuration & $\times$ \\
  Memory read \& write permission & $\times$ \\ \hline \vspace{-3em}
\end{tabular}}
\end{table}

\section{DNN-Defender Mechanism} 
The DNN-Defender mechanism is developed to utilize the minimum number of swap operations and to optimize latency overheads to protect the DRAM against the targeted RH in the adversarial weight attack. Our mechanism makes traceability very difficult for the attacker. Even though the attacker can precisely locate the target data, the DNN-Defender only requires to perform multiple swap operations to the victim rows to secure the memory.  
The memory instruction-based swap allows us to avoid concerns regarding invalid refreshes resulting from a communication delay between the counter and the counter table and off-chip access as seen in previous designs \cite{seyedzadeh2016counter,woo2022scalable,saileshwar2022randomized,lee2019twice,park2020graphene}. Besides, our design prioritizes versatility over assessing the potential impact of swap operations on data in different scenarios and determining the worst-case scenario. The focus is on preventing all threats regardless of the circumstances. However, since an aggressor row is typically accompanied by more than one victim row, one can inevitably ignore the other victim rows when focusing on the target data since the impact of such data on the final result is far less than that of important data. However, to minimize the adverse effect, DNN-Defender also uses the lowest resources possible to ensure their security.

\begin{figure}[t]
\begin{center}
\begin{tabular}{c}
\includegraphics [width=0.89\linewidth]{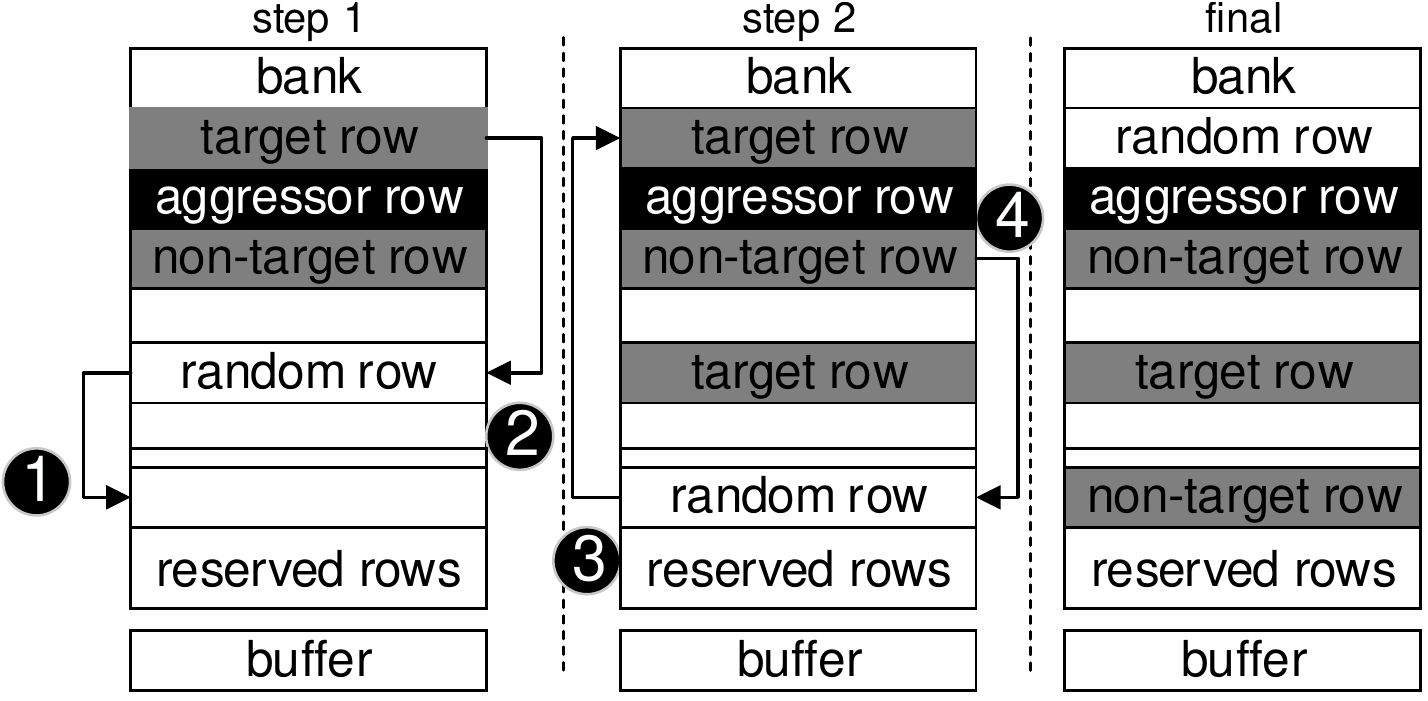}\vspace{-0.4em}
 \end{tabular} \vspace{-0.9em}
\caption{The four-step swap mechanism in DNN-Defender.}\vspace{-1em}
\label{DDSWAP}\vspace{-1.0em}
\end{center}
\end{figure}

As shown in Fig. \ref{DDSWAP}, we propose to virtually partition the protected data region in each memory sub-array into the target and non-target victim rows according to their protection priority. The target row holding the targeted DNN weight is the row with the highest priority to be protected. In other words, if a bit(s) within a target row is flipped, the final DNN accuracy will significantly drop. The non-target rows however show a certain degree of tolerance to errors as they have no/negligible effect on the final result as illustrated in Fig. \ref{RHthre}(b). However, in extreme continuous attack scenarios, it is not ruled out that non-target rows may contain important data that would adversely reduce the DNN accuracy. Therefore, as illustrated in Algorithm \ref{alg}, DNN-Defender not only needs to guarantee the security of the target row but also furnishes a low-cost safeguard for the non-target row.
The DNN-Defender's swap operations are accomplished in four steps. As depicted in Fig. \ref{DDSWAP}, in step \encircle{1}, the memory controller selects a random row in the sub-array and leverages RowClone \cite{seshadri2013rowclone} to copy it to the reserved row. In step \encircle{2}, the target row is copied to the random row in the same way. This in-memory operation will refresh the target row and reset the attacker's target. The rationale is that even though the target row is copied to another position, the malicious process knows the new location, so it will move to the latest row beside the swapped target row and make it a new aggressor row.
In step \encircle{3}, the random row in the reserved rows region is copied back to the original location of the target row. Aiming to refresh the non-target row knowing that the capacity of the reserved rows region is limited, in step \encircle{4}, the non-target row is copied to the reserved row until the next random row overwrites it. Finally, the original target row and random row are swapped, and the non-target row is refreshed. Please note that the attacker can track the target rows and attack the updated addresses, while they will no longer attack the non-target and random rows.
It is worth pointing out 
unlike SRS \cite{woo2022scalable}  and RRS \cite{saileshwar2022randomized}, which focus on the aggressor row, DNN-Defender focuses on protecting the victim row.

 \begin{algorithm}[b]
        \caption{\small DNN-Defender's Swap Algorithm}
          \scalebox{0.62}{
    \begin{minipage}{1.8\linewidth}
        \begin{algorithmic}[1]
            \State $\textbf{Procedure: \textit{Protection}}$
                \State $\textbf{If} \hspace{4pt} DD\_Start$ 
                \State $\hspace{16pt}Define \ \ random\_row = rand(DRAM)$ 
                \State $\hspace{16pt}reserved\_row \gets random\_row$ 
                \State $\hspace{16pt}random\_row \gets targe\_row[row]$ 
                \State $\hspace{16pt}targe\_row[row] \gets reserved\_row$
                \State $\hspace{32pt}\textbf{If}\hspace{4pt}(Target\_rows == 1)\hspace{4pt}break;$
                \State  $\hspace{32pt} \textbf{else} \hspace{4pt}
                        \textbf{For} \hspace{4pt} row \hspace{4pt} in \hspace{4pt} Target\_rows \hspace{4pt} \textbf{do}$
                            \State $\hspace{48pt}reserved\_row \gets non\_target\_row[row];$ 
                            \State $\hspace{48pt}non\_target\_row[row] \gets targe\_row[row+1];$ 
                            \State $\hspace{48pt}targe\_row[row+1] \gets non\_target\_row[row];$ 
                \State $\textbf{else} \hspace{4pt}
                    \textbf{if}\hspace{4pt} DD\_Interrup$
                        \State $\hspace{16pt}break;$
                    \State $\textbf{else}
                        \hspace{4pt} continue();$

    \State \textbf{end} $\textbf{Procedure}$
        \end{algorithmic} 
         \end{minipage} \vspace{-3em}}
         \label{alg}
    \end{algorithm}

\begin{figure}[t]
\begin{center}
\begin{tabular}{c}
\includegraphics [width=0.70\linewidth]{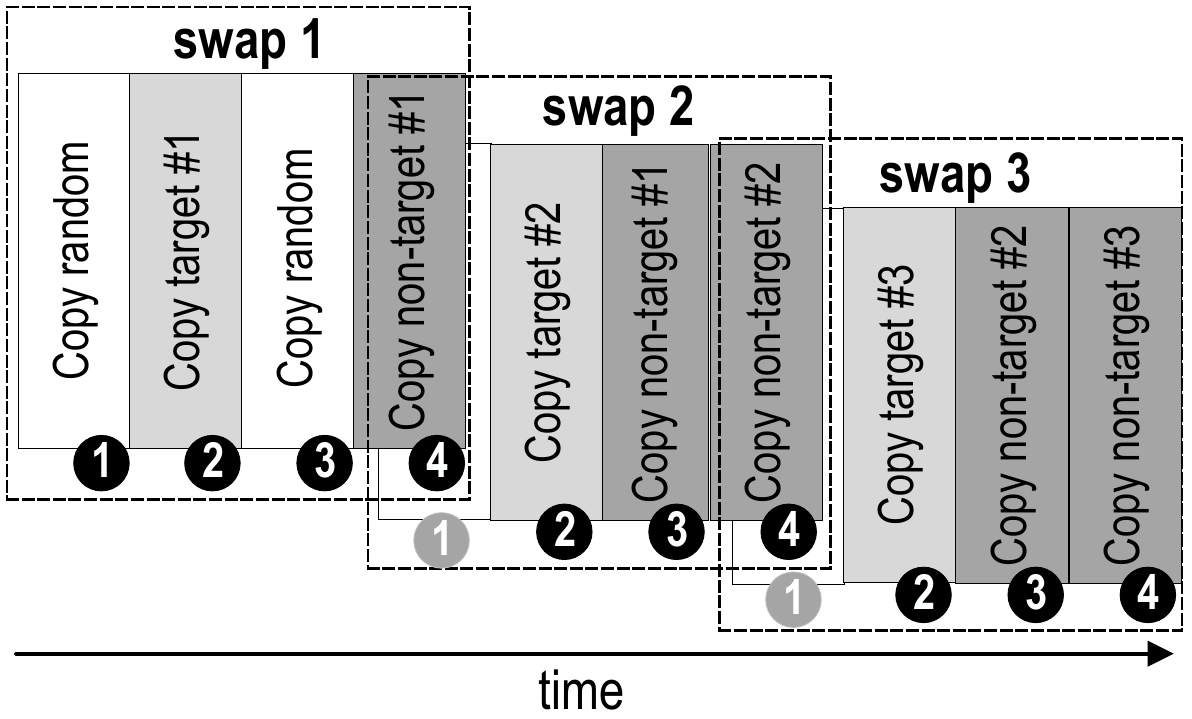}\vspace{-0.4em}
 \end{tabular} \vspace{-1em}
\caption{Timeline of a sample DNN with multiple swap ops.}\vspace{-1.0em}
\label{swapflow}\vspace{-1em}
\end{center}
\end{figure}

\textbf{Timing Considerations.} Considering $T_{RH}$ is set to 4,800 in LPDDR4 \cite{woo2022scalable}, the victim rows must be refreshed before the activation number reaches the threshold. Therefore, it is necessary for DNN-Defender to complete swapping operations within the threshold window ($<$4800)$\times T_{ACT}$. 
DNN-Defender only needs one random row generation to support all swap operations. 
Figure \ref{swapflow} illustrates the DNN-Defender's defense timeline and parallelism for a sample DNN with multiple swap operations. In swap 1, it requires a Random Number Generator (RNG) to define the initial random row for step \encircle{1}. The remaining three steps then follow that as we discussed before in Fig. \ref{DDSWAP}. After completing step \encircle{4}, DNN-Defender stores the non-target row data of swap 1 in the reserved row, which means that the non-target row 1 at this time can be used as a new random row. Therefore, as shown in Fig. \ref{swapflow}, step \encircle{1} of swap 2 can overlap with step \encircle{4} of swap 1. Then it finishes the remaining three steps (\encircle{2} to \encircle{4}). This method is readily applied to the next swaps.

\textbf{Priority Protection Mechanism.}
To select the target rows requiring high-priority protection from DNN-Defender, we propose using the same attack searching algorithm adopted by an attacker for BFA~\cite{rakin2019bit}. We use a copy of the victim model with identical model architecture and weights to compute the gradient of the loss function w.r.t. each weight bit ($ |\nabla_{\tB_l} \mathcal{L}|$). Similar to the BFA attack, we rank the gradients and only flip the highest gradient bit of a specific layer which causes the largest increase in loss function as shown in Eqn. \ref{eqt:BFAN}. We run the software search algorithms until the model accuracy drops close to the random guess level (e.g., 10 \% for CIFAR-10). After performing one complete round of BFA, we record the target bit location $R_1 \in (l,k)$ ($l$ is the layer number, $k$ is the index at a specific layer) that was flipped in the current attack round. Next, we flip back all the targeted bits and perform another round of BFA but this time skip flipping the bits from the previous round $R_1$. Hence for round $R_2$, if we encounter any bits from the previous round $R_1$, we skip it and select the next bit candidate based on gradient. In this way, we keep performing the bit-search algorithm for multiple rounds ($R_c,c=1,2,...,r$), where each round skips all the bits from the previous rounds of the attack. The total number of rounds $r$ depends on the number of bits the defender wants to secure, and increasing $r$ will increase the level of protection. After profiling all the vulnerable bits to BFA for multiple rounds, we select these bit sets as the priority bits requiring more protection. DNN-Defender will prioritize protecting these vulnerable bit sets by selecting them as the target row in their corresponding sub-array. \vspace{-1em}

\section{Experimental Results}
\begin{figure}[t]
\begin{center}
\begin{tabular}{c}
\includegraphics [width=0.99\linewidth]{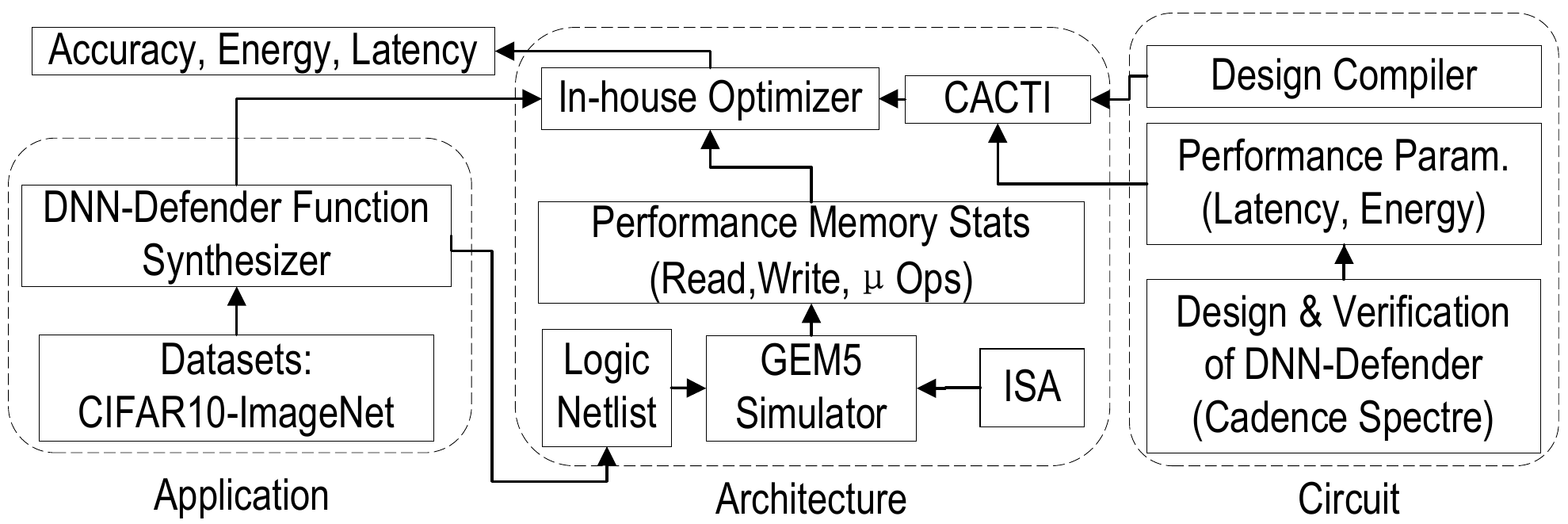}\vspace{-0.5em}
 \end{tabular} \vspace{-1.4em}
\caption{Proposed evaluation framework.}\vspace{-1.5em}
\label{flowchart}
\end{center}
\end{figure}

\textbf{Setup.} We present a cross-layer evaluation framework as shown in Fig. \ref{flowchart} to demonstrate the benefits of DNN-Defender. Firstly, we developed DNN-Defender's sub-arrays with peripherals using Cadence Spectre in the 45nm NCSU PDK library \cite{NCSU_PDK} at the circuit-level to verify functionality, attain performance parameters, and measure the row-shuffle time. The memory controller and registers were designed and synthesized by Design Compiler with a 45nm industry library. Afterward, we incorporated the results from circuit-level assessments and extensively modified CACTI at the architecture-level. We implemented DNN-Defender's ISA in gem5 \cite{binkert2011gem5}, and exported the memory statistics and performance to an in-house C++ DNN-Defender optimizer, taking the CACTI output and application netlist as the inputs. At the application, we evaluated the performance of our proposed technique in defending against adversarial BFA using two commonly-used visual datasets: CIFAR-10 and ImageNet. The weights were quantized to 8-bit width. To carry out the BFA, we randomly sampled images from the test/validation set, with a default sample size of 128 for both datasets. \vspace{-0.5em}

\begin{table}[b] \vspace{-1em}
\caption{Comparison with prior generic RowHammar mitigation frameworks.} 
\begin{center}\vspace{-1em}
\scalebox{0.7}{
\begin{tabular}{ccccc}
\hline
Framework                                                  & involved memory & capacity overhead & area overhead  \\ \hline
Graphene \cite{park2020graphene}          & CAM-SRAM        & 0.53MB$^\ddagger$+1.12MB$^\dagger$      & 1 counter      \\
Hydra \cite{qureshi2022hydra}             & SRAM-DRAM       & 56KB$^\dagger$+4MB$^*$          & 1 counter      \\
TWiCE \cite{lee2019twice}                 & SRAM-CAM        & 3.16MB$^\dagger$+1.6MB$^\ddagger$      & 1 counter      \\
Counter per Row                                            & DRAM            & 32MB$^*$              & 16384 counters \\
Counter Tree \cite{seyedzadeh2016counter} & DRAM            & 2MB$^*$               & 1024 counters  \\
RRS \cite{saileshwar2022randomized}                & DRAM-SRAM      & 4MB$^*$+NR$^\dagger$          & NULL           \\
SRS \cite{woo2022scalable}                & DRAM-SRAM      & 1.26MB$^*$+NR$^\dagger$        & NULL           \\
SHADOW \cite{wi2023shadow}                & DRAM            & 0.16MB$^*$            & 0.6\%          \\
P-PIM \cite{ranyang2023ppim}                 & DRAM            & 4.125MB$^*$           & 0.34\%         \\
\emph{\textbf{DNN-Defender}}                               & \emph{DRAM}            & \emph{\textbf{0}}                 & \emph{\textbf{0.02\%}}           \\ \hline
\end{tabular}}
\end{center} 
\label{eva}

\tiny NR = Not Reported\\
$^*$\tiny The capacity overhead of DRAM.
$^\dagger$\tiny The capacity overhead of SRAM. 
$^\ddagger$\tiny The capacity overhead of CAM. \vspace{-2em}
\end{table} 

\subsection{Performance Evaluation}
\textbf{Hardware Overhead Analysis.}
We compare the DNN-Defender's hardware overhead with the latest generic RH mitigation mechanisms in the literature in Table \ref{eva}.
For this experiment, our strategy is to consider the same 32GB: 16-bank of DDR4 DRAM for all frameworks and normalize the capacity overhead and area overhead across different frameworks. In Table \ref{eva}, 
\textit{i)} the involved memory indicates the type of memory used by the framework for RH protection. As discussed, certain frameworks rely on a counter to monitor intrusions and store tracking information in the system using CAM/SRAM. Nonetheless, because of the considerably higher cost of CAM and SRAM in comparison to DRAM, selecting a framework with such supplementary resources may be controversial. For example, Table \ref{eva} shows that only Graphene \cite{park2020graphene} and TWiCE \cite{lee2019twice} occupy two fast storage resources simultaneously.
At the same time, Hydra \cite{qureshi2022hydra}, SRS \cite{woo2022scalable}, and RRS \cite{saileshwar2022randomized} rely on SRAM in addition to DRAM. 
\textit{ii)} Capacity overhead refers to the memory resources utilized by RH in a framework. Such resources are dedicated solely to RH and not for other purposes. Take Graphene \cite{park2020graphene} as an example, it requires storing counting tables in SRAM (1.12 MB), and the space occupied by these tables can no longer be used as a shared storage space to store data. Graphene also requires 0.53 MB CAM to track vulnerable rows. Take SHADOW \cite{wi2023shadow} as an example, 0.16 MB capacity of DRAM is dedicated to enabling RH mitigation.
Our framework stands out from others as it does not utilize any additional memory resources for RH mitigation. Unlike Graphene \cite{park2020graphene}, TWiCE \cite{lee2019twice}, SRS \cite{woo2022scalable}, and RRS \cite{saileshwar2022randomized}, DNN-Defender does not require any fast-read memory. Furthermore, in contrast to SHADOW \cite{wi2023shadow} and P-PIM \cite{ranyang2023ppim}, DNN-Defender does not even sacrifice DRAM resources, where all rows can be used for storing data in the same way as ordinary rows. \textit{iii)} Some frameworks not only require storage device resources but also additional components for RH mitigation. We can find that DNN-Defender offers one of the most area-efficient solutions compared to other frameworks. 

\begin{figure}[t]
\begin{center}
\begin{tabular}{c}
\includegraphics [width=1.01\linewidth]{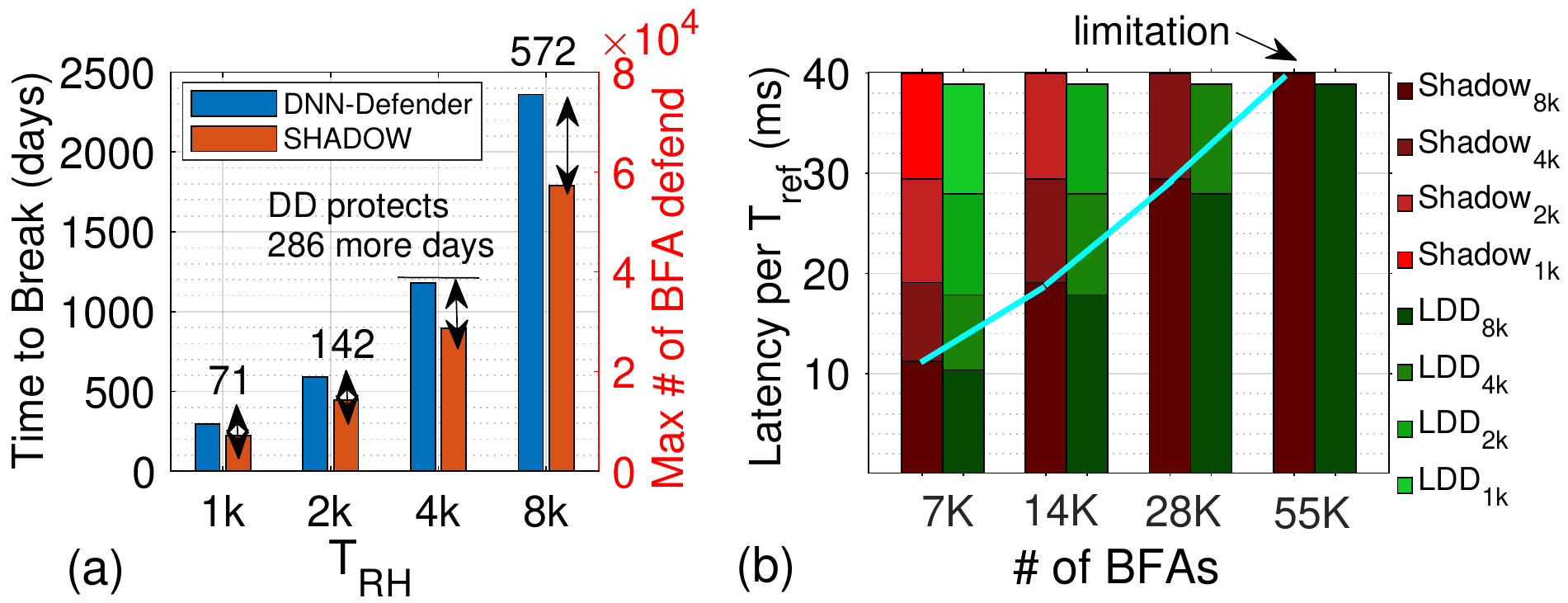}\vspace{-0.5em}
 \end{tabular} \vspace{-0.7em}
\caption{(a) Time-to-break DNN-Defender (DD) and SHADOW in different RowHammer thresholds, (b) Latency of DNN-Defender and SHADOW \cite{wi2023shadow} at different no. of BFA.}\vspace{-2em}
\label{security}
\end{center}
\end{figure}

\textbf{Security \& Performance Analysis.}
We establish that the system's security level is directly proportional to the time taken by an attacker to breach it, i.e., the longer it takes to breach the system, the higher the security level. The first assumption is that the data has no unique mapping, so we consider the vulnerable data rows to be evenly distributed in all banks. Therefore, the number of data rows we need to protect in each bank is given by \textit{$N_s=S_{bit}/banks$}, where $N_s$ is finite. In other words, when the number of under-attack rows increases, $N_s$ will exceed the defendable threshold. It is, therefore, essential to identify the threshold that our framework can withstand. 
Assuming the worst-case scenario in which a row contains only a single target weight bit, the maximum number of defended BFA corresponds to the number of target rows. As DNN-Defender doesn't modify the circuitry of the original DRAM array, we can utilize the timing baseline of DRAM without any alterations as given in \cite{wi2023shadow} to calculate the time required for swap operation, where $T_{swap}=3\times T_{AAP}$, $T_{AAP}$ = 90ns.
As discussed, $T_{RH}$ depends on the DRAM process node as the minimum number of activations that can impose bit-flip.
So the time constraint for DNN-Defender to perform the swap operations is given by $T_{ACT}\times T_{RH}$, and the maximum number of swap operations can be calculated as $\frac{T_{ACT}\times T_{RH}} {T_{swap}}$.
We can calculate $T_n=T_{ACT}\times T_{RH} + T_{swap}\times N_s$, and then find the total number of swap operations in single $T_{ref}$ by $N=\frac{T_{ref}}{T_n}\times N_s$.
Our experiments show that DNN-Defender and SHADOW \cite{wi2023shadow} are the \ul{\textit{only frameworks that can withstand the white-box attacks}}, therefore, we only report their security level analysis. Our results indicate that even the SRS mechanism \cite{woo2022scalable} cannot defend against white-box attacks for a period of one day. Figure \ref{security}(a) reports the time-to-break in days for various $T_{RH}$s and the corresponding number of the defended BFAs. We observe that our framework outperforms SHADOW in 1k, 2k, 4k, and 8k thresholds, e.g., as indicated for $T_{RH}$=4k, the attacker will require $\sim$1180 days to break a DNN-Defender-supported system, while SHADOW stands up for $\sim$894 days. 

Figure \ref{security}(b) shows DNN-Defender and SHADOW \cite{wi2023shadow} with 1k, 2k, 4k, and 8k RH thresholds. As mentioned earlier, we pick SHADOW as one of the best RH mitigation mechanisms over others for this comparison.
The figure illustrates that with an increase in the number of BFAs, the rate of latency increase decelerates and eventually reaches a limit for both frameworks. We chose four critical points of BFAs, i.e., 7K, 14K, 28K, and 55K, where each corresponds to the maximum allowable number of BFAs under various thresholds. This selection is made to facilitate a comparative analysis. Considering a 4k threshold, an escalation in the number of BFAs leads to a peak in latency. When compared to SHADOW operating at the same threshold, our framework exhibits lower latency in all cases.
We consider the power consumption of a standard DRAM process as a benchmark to fully reveal all aspects of our proposed framework. Compared with other frameworks, DNN-Defender indicates no significant power-saving. For instance, even when SHADOW is set with a threshold of 1k, DNN-Defender shows a negligible 1.6\% power-saving. However, considering the power consumption by SRAM-based frameworks such as SRS and RRS and off-chip memory communication in such systems, DNN-Defender offers a significant improvement (3.4$\times$ compared with SRS). \vspace{-0.8em}

\subsection{Defense Evaluation}
\noindent\textbf{Evaluation of DNN-Defender against Semi-White-Box BFA.} First, we consider a naive BFA attack~\cite{rakin2019bit} where the attacker is not aware of our defense strategy (Semi-White-box). Such a naive attack will fail since the targeted bit-flip sequence is not optimized to bypass our defense. A naive attacker will generate a small sequence of target bits and will attempt to flip them. However, our defense will eliminate the impact of bit-flips via the swap operation, and the attacker will not achieve any success (i.e., accuracy degradation) using the existing BFA algorithm. 

\noindent\textbf{Evaluation of DNN-Defender against White-Box BFA.}
With complete white-box knowledge, the attacker is aware of our defense and tries to evade it through adaptive search. To bypass DNN-Defender, the attacker can generate multiple sequences of targeted bits offline using a copy of the target model and evaluate the attack's success by launching the attack to the victim space. However, if a specific chain of bit sequences fails to degrade model accuracy, the attacker can skip this sequence to generate a new set of bit sequences for the next attack round. This way, the attacker will iterate through all the Secured Bits (SB) protected by DNN-Defender and still observe no success i.e., accuracy drop at the output.

\begin{figure}[t]
\begin{center}
\begin{tabular}{c}
\includegraphics [width=1\linewidth]{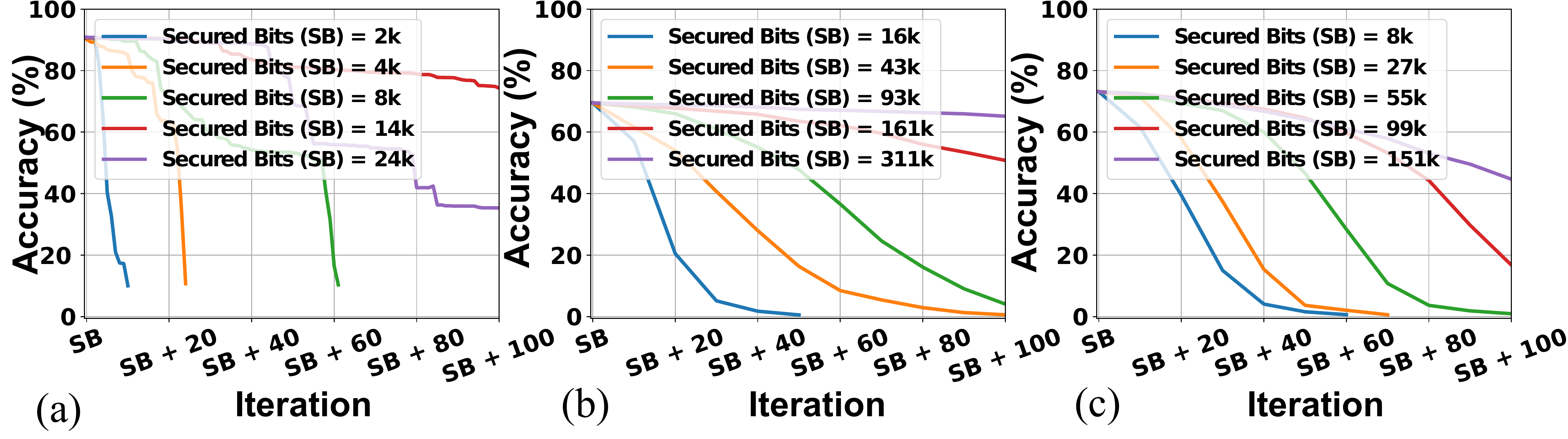}\vspace{-1.0em}
 \end{tabular} \vspace{-0.2em}
\caption{DNN-Defender evaluation for different amounts of Secured Bits (SB) for (a) VGG-11 trained on CIFAR-10, (b) ResNet-18 trained on Imagenet, and (c) ResNet-34 trained on Imagenet against BFA.}\vspace{-1.8em}
\label{fig:cifar-imagenet-results}
\end{center}
\end{figure}

Figure \ref{fig:cifar-imagenet-results} demonstrates the effectiveness of DNN-Defender in mitigating the performance degradation caused by BFA when the attacker further adapts the search and attempts to flip more bits than protected by our defense (denoted as SB + \# of additional bit-flips). The plots show multiple curves of performance degradation on each of the evaluation models (e.g., VGG-11, ResNet-18, ResNet-34), with each curve representing the degradation after securing a specific number of bits using DNN-Defender denoted as Secured Bits (SB). The plots reveal that as the SB increases, it takes the attacker an increasing number of iterations to cause the same performance degradation, gradually deteriorating the performance of BFA near the random attack level. For example, in Fig.~\ref{fig:cifar-imagenet-results}, for the VGG-11 model, increasing the secured bits from 2k to 8k increases the number of additional bit-flips required to achieve the same attack efficacy by $\sim 6 \times$. Eventually, even this adaptive white-box BFA attack loses its potency after securing 24k bits (still only $\sim$ 4\% of the total model bits). In conclusion, our proposed DNN-Defender will increase the attack time and effort by requiring an increasing amount of bit-flips to achieve the same attack efficacy as the baseline (no defense). If we secure a large number of bits (e.g., $\sim$ 24k for VGG-11), then even after increasing the bit-flip, the effect of BFA can be reduced to a random attack level. 

\begin{table}[b]
\centering
\caption{{Comparison to BFA software defences on CIFAR-10 evaluated attacking a ResNet-20 model.}} \vspace{-1em}
\label{tab:cmp}
\scalebox{0.6}{
\begin{tabular}{@{}cccc@{}}
\toprule
\begin{tabular}[c]{@{}c@{}} {Models} \end{tabular} & \begin{tabular}[c]{@{}c@{}}{Clean Acc.}(\%)\end{tabular} & \begin{tabular}[c]{@{}c@{}}{Post-Attack acc.}(\%)\end{tabular} & \begin{tabular}[c]{@{}c@{}} {Bit-Flips \#} \end{tabular} \\ \midrule
Baseline ResNet-20~\cite{rakin2019bit}  & 91.71 & 10.90 & 20 \\
Piece-wise Clustering~\cite{he2020defending}  &  90.02& 10.09 & 42 \\
Binary weight~\cite{he2020defending} & 89.01 & 10.99 & 89 \\
Model Capacity $\times$ 16~\cite{rakin2021ra} & 93.7 & 10.00 & 49 \\
Weight Reconstruction~\cite{li2020defending}  & 88.79 & 10.00 & 79  \\
RA-BNN~\cite{rakin2021ra}  & 90.18 & 10.00 & 1150 \\ 
RRS~\cite{saileshwar2022randomized}  & 91.71 & 75.65 & 342 \\ 
SRS~\cite{woo2022scalable}  & 91.71 & 75.92 & 378 \\ 
SHADOW~\cite{wi2023shadow}  & 91.71 & 88.80 & 985 \\ 
\textbf{\emph{DNN-Defender}}  & \textbf{\emph{91.71}} & \textbf{\emph{ 91.71}} & \textbf{\emph{1150}} \\  \bottomrule
\end{tabular}} \vspace{-1.5em}
\end{table}

\textbf{Comparison to BFA Defenses.}
In Table~\ref{tab:cmp}, we compare DNN-Defender against the existing \textit{training-based DNN software defenses}~\cite{rakin2021ra,he2020defending,li2020defending} and  \textit{selected generic hardware defenses}, i.e., RRS \cite{saileshwar2022randomized}, SRS \cite{woo2022scalable}, and SHADOW \cite{wi2023shadow}. A general approach among the prior software-based defense works is to reduce model weight bit-width precision~\cite{hong2019terminal,he2020defending} and increase the model size to reduce the impact of weight noise on accuracy~\cite{rakin2021ra}. Here, a binary neural network~\cite{rakin2021ra} with both binary weight and activation achieves the best defense performance against the BFA, requiring over 1000 bit-flip to reduce the model accuracy close to random guesses. When protecting the exact required number of (e.g., 1150) vulnerable bits, DNN-Defender can resist the BFA attack better than the binary model. However, our attack incurs slight hardware (e.g., latency \& energy) overhead which is not the case for software training algorithms~\cite{rakin2021ra,he2020defending,li2020defending}.
In contrast, all the software-based training methods suffer from high training overhead and model accuracy drop. Our method is an effective defense against BFA w/o requiring any training overhead or performance drop with minimal hardware overhead. Additionally, any software protection~\cite{li2021radar} or DNN training algorithm~\cite{he2020defending,rakin2021ra} is not necessarily a competing method against our defense; prior training-based defenses can further boost the protection against BFA on top of our method. In addition, we observe that DNN-Defender shows higher post-attack accuracy compared with prior designs withstanding more number of BFAs.
\vspace{-1em}

\section{conclusions}
Herein, we proposed a powerful DRAM-based defense mechanism called DNN-Defender that leverages the potential of in-DRAM swapping to protect quantized DNNs from targeted RowHammer bit-flip attacks.
Our results indicate that DNN-Defender is capable of providing robust protection against consecutive targeted RowHammer attacks on CIFAR-10 and ImageNet datasets downgrading its performance to a random attack level. \vspace{-1em}

\section*{Acknowledgment}
\small This work is supported in part by the National Science Foundation (NSF) under grant no. 2228028.\vspace{-1em}

\bibliographystyle{ACM-Reference-Format}
\bibliography{References}\vspace{-2em}

\end{document}